\documentclass[prl,aps,twocolumn,showpacs]{revtex4}
\usepackage{epsfig}
\usepackage{graphicx}
\begin{document}
\input{epsf}
\title{Dynamics and pattern formation in invasive tumor growth}
\author{Evgeniy Khain and Leonard M. Sander}
\affiliation{Department of Physics and Michigan Center for
Theoretical Physics, The University of Michigan, Ann Arbor,
Michigan 48109}

\begin{abstract}
In this work, we study the in-vitro dynamics of the most malignant
form of the primary brain tumor: Glioblastoma Multiforme.
Typically, the growing tumor consists of the inner dense
proliferating zone and the outer less dense invasive region.
Experiments with different types of cells show
qualitatively different behavior. Wild-type
cells invade a spherically symmetric manner, but mutant
cells are organized in tenuous branches. We formulate a model for
this sort of growth using two coupled reaction-diffusion equations
for the cell and nutrient concentrations. When
the ratio of the nutrient and cell diffusion coefficients exceeds
some critical value, the plane propagating front becomes unstable
with respect to transversal perturbations. The instability
threshold and the full phase-plane diagram in the parameter space
are determined. The results are in a good agreement with
experimental findings for the two types of cells.
\end{abstract}
 \maketitle

\begin{figure}[ht]
\vspace{-0.5cm}
\centerline{\includegraphics[width=5cm,clip=]{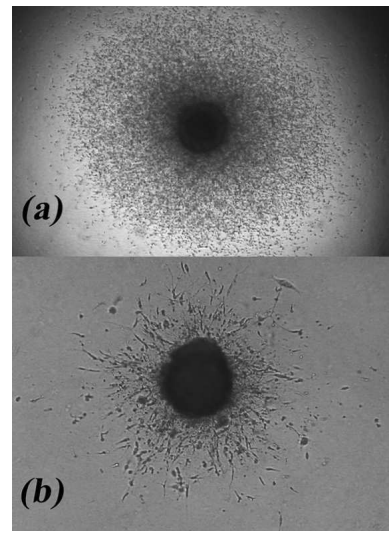}}
\caption{Growing tumors from in {\it
vitro} experiments \cite{Stein} in collagen gel for the wild-type
(a) and mutant (b) cells. These \emph{in vitro} tumors consist of an
inner proliferation zone with a very high density of cells and an
outer invasive zone, where the cell density is smaller. The structure 
\emph{in vivo} is believed to be similar.
The radius of the inner zone here is about 250 $\mu$.
For wild-type cells a spherically symmetric pattern
is observed, (a).  Mutant cells are organized in tenuous
branches, (b). Note also that the invasive region for the mutant
type cells grows slower than for wild-type cells. } 
\label{tumor}
\end{figure}

One of the most aggressive forms of primary
brain tumor is Glioblastoma Multiforme (GBM) \cite{Surawicz}.
Despite major advances in medical science the  prognosis for victims of this disease is very poor \cite{Surawicz}: the median survival for patients with newly
diagnosed GBM is approximately 12 months. One of the main
reasons for such high mortality and poor success of 
treatment is the fact that GBMs are highly invasive
\cite{Demuth2}. The growing tumor sheds invasive cells which run through the brain, see Fig.~\ref{tumor}. The invasive
nature of malignant gliomas  makes treatment 
difficult  \cite{Demuth2};  secondary tumors are produced by the invasive cells even if the primary is removed. In this paper we introduce a reaction-diffusion model for invasion. By comparing two different cell lines we hope to get insight into invasion dynamics which is needs to be better  understood.

This work is inspired by recent \emph{in vitro} experiments \cite{Deisboeck,Stein}  where 
microscopic tumor spheroids (radius about 250 $\mu$) were placed
in collagen-I gel and allowed to grow. The cell lines used were U87 and
U87-$\Delta$EGFR. The first type is called `wild-type' in what follows. 
The second is a mutant line \cite{Nagane} in which there is an amplification of the epidermal growth factor receptor (EGFR) gene. This amplification occurs
in approximately $40$ percent of cases of GBM \cite{Huncharek}.

If we compare the growth of the two cell lines \emph{in vitro} we see two main differences; cf. Figure~\ref{tumor}. The invasive
region for the wild-type cells grows faster than for the mutant
cells, and the wild-type  produces a
spherically symmetric pattern, whereas mutant
cells produce a branching pattern \cite{Stein}. 
In the present work, we formulate a simple reaction-diffusion model that
is able to reproduce these experimental findings and may give insight 
into the functional significance of the mutation.

We interpret the  branching shown in Fig.~\ref{tumor}b,  as a
branching instability. Analogous instabilities were studied in the theory of combustion
\cite{Sivashinsky}, and in studies of the self-organization
of microorganisms \cite{Ben-Jacob}. One way of modeling this
is to assume that there is attraction between
cells \cite{Sander} due to the production of growth factors.
Another possibility is to assume nonlinear
diffusion, where the diffusion coefficient {\it increases} with
the density of the cells, as was proposed for  bacterial
colonies \cite{Ben-Jacob}. We will introduce another mechanism
based on the known biology of GBM.

In our continuum description, we will deal with the density of
cells $u(\mathbf{r},t)$ and the density of some growth factor
or nutrient (whichever controls the growth), $c(\mathbf{r},t).$
We assume that $c$ diffuses to the tumor from far away.
Each cancer cell is able to proliferate as well as to perform
random motion. It is known that within the inner  region, cells
have quite a high proliferation rate whereas in the invasive region, cells have high motility, but low proliferation rate  \cite{Giese}. This is an
indication of a dynamical switch between the cell phenotypes
\cite{Giese}. We model it by introducing a 
\emph{density-dependent} proliferation term where the proliferation rate increases
with cell density. The simplest form for such a term is $\partial u/\partial t \propto  u^2 c$. (The 
extra power of $u$ compared to the usual $uc$ means that high density 
gives rapid proliferation since the proliferation rate per cell, $(1/u)\partial u/\partial t \propto u$.) As we will see, 
this term drives the instability and leads to branching. 
We assume that in order to proliferate a cell needs to consume some
amount of $c$. The density $c$ obeys a diffusion equation with a
sink at the tumor cells. 

We encode these assumptions in the following equations:
\begin{eqnarray}
&&\frac{\partial u}{\partial t} = \mathbf{\nabla} \cdot ( D_u \,
\mathbf{\nabla} u) + \alpha \, u^2 c \,, \nonumber \\
&&\frac{\partial c}{\partial t} =  \mathbf{\nabla} \cdot ( D_c \,
\mathbf{\nabla} c) - \beta \, u^2 c\,. \label{basic}
\end{eqnarray}
Here $D_u$ and $D_c$ are the diffusion coefficients of $u$ and
$c$, $\alpha$ is a proliferation
coefficient, and $\beta$ is the coefficient of nutrient
consumption. We assume that the density in the center of the tumor
is not very high compared to the density of closely packed cells
$u_c$. We suppose also that the nutrient concentration is kept
constant far from the tumor: $\lim_{r \rightarrow \infty} c(r) =
c_\infty$. 

In what follows we will measure cell
density in the units of some characteristic density $u_0$,
nutrient density in units of $c_\infty$, distance in units of
$[D_c/(\beta {u_0}^2)]^{1/2}$, and time in units of $(\beta
{u_0}^2)^{-1}$. This gives:
\begin{eqnarray}
&&\frac{\partial u}{\partial t} =
\frac{1}{\delta}\mathbf{\nabla}^2
u + \frac{1}{m}\, u^2\,c \,, \nonumber \\
&&\frac{\partial c}{\partial t} =  \mathbf{\nabla}^2 c - u^2\,
c\,, \label{basicdim}
\end{eqnarray}
where $\delta = D_c/D_u$ is the ratio of the nutrient and cell
diffusion coefficients and $m = \beta u_0/(\alpha c_\infty)$ is
the ratio of consumption and proliferation rates. 

Typically, the nutrient or growth factor represented by $c$ is a small molecule. It 
is expected to diffuse much faster than the cells. For example,  the diffusion
coefficient of glucose in the brain is of the order of $10^{-7}
\mbox{cm}^2/\mbox{s}$, while the cell diffusion is of the order of
$10^{-9} \mbox{cm}^2/\mbox{s}$ \cite{Sander}, so that $\delta \sim 100$. A typical
nutrient consumption is $10^{-12} \mbox{g/cell/min}$ \cite{Li},
and a typical glucose concentration is of the order of $1
\mbox{g/l}$. Assuming that typical cell density within the
invasive region is of the order of $10^5 \mbox{cell}/\mbox{cm}^3$,
we estimate the consumption rate as $1.7\times 10^{-6}
\mbox{s}^{-1}$. The typical proliferation rate in experiments
\cite{Stein} is of the order of $1/\mbox{day}$, so that $m$ turns
out to be of the order of $0.1$. 

We work  in a
two-dimensional channel geometry. Let $x$ be the direction of tumor growth, and
$y$ be the transverse direction, perpendicular to the direction of
the front propagation. In the $y$ direction we use periodic boundary conditions
with a finite channel width. In the $x$ direction,  far ahead of  the tumor, the cell concentration
is zero, $u(x=\infty) = 0$, and the scaled nutrient concentration
is unity, $c(x = \infty) = 1$. On the other hand, at $x = -\infty$
we demand $c = 0$. There is a conservation law in
Eqs.~(\ref{basicdim}): a volume integral over the system of $(m u
+ c)$ is a conserved quantity. Its interpretation is that in our model a cell needs some amount of food to divide.
Therefore, at $x = -\infty$, $u = 1/m$ if there is a steady state. Our initial conditions are 
$u=1/m$ for  $x\le 0$; $u=0$ for $x >0$, and $c=0$,  for $x\le 0$; $c=c_\infty$ for $x>0$. We will investigate the situation after transients have died away and a steady propagating state has been established.

\begin{figure}[ht]
\vspace{-0.4cm}
\centerline{\includegraphics[width=6.5cm,clip=]{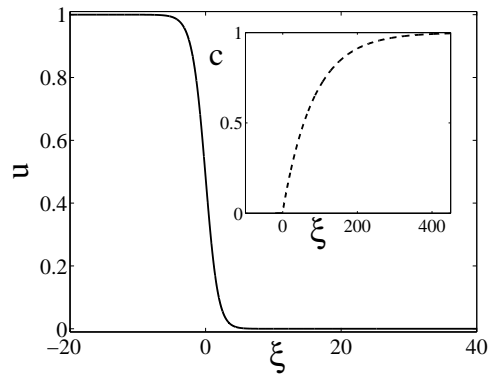}}
\caption{Density profiles of cells and
nutrient from Eqs.~(\ref{plane}). The first
curve is the cell density $u=u(\xi)$ (solid line.) 
The second curve (dashed) is $c=c(\xi)$ (inset). 
 The parameters are $m=1$, $\delta = 100$.}
\label{profile1}
\end{figure}

First, we consider the solutions of Eqs.~(\ref{basicdim}) in the
form of plane propagating fronts: $u = u_0(\xi); c = c_0(\xi)$,
$\xi = x-vt$. Substituting into Eqs.~(\ref{basicdim}) we arrive
at:
\begin{eqnarray}
&& \frac{1}{\delta}{u_0}^{\prime\prime} + v {{u_0}^\prime}
+ \frac{1}{m}\, {u_0}^2\,c_0 = 0 \,, \nonumber \\
&&{c_0}^{\prime\prime} + v {{c_0}^\prime} -{u_0}^2\,c_0 = 0\,.
\label{plane}
\end{eqnarray}
To obtain the profiles and the velocity of front propagation, we
performed (in Matlab) the following shooting procedure. First, we
write down Eqs.~(\ref{plane}) as four coupled first-order
differential equations in the form $(\vec{a})^{\prime} = M
(\vec{a})$, where $\vec{a}$ is the column of solutions with
elements ${u_0}, {u_0}^{\prime}, {c_0}, {c_0}^{\prime}$. Then we
find the eigenvalues and eigenvectors of $M$ at $\xi = \pm\infty$.
Starting with the solution $\vec{a}$ that is proportional to the
eigenvector belonging to the positive eigenvalue at $\xi = -
\infty$ (using the linearity of the problem, we choose the
constant of proportionality to be unity), we perform shooting by
the velocity of front propagation $v$. We find the profiles by
demanding that the solution $\vec{a}$ at $\xi = + \infty$ is a
linear combination of the two eigenvectors $\vec{\psi}_1$ and
$\vec{\psi}_2$ belonging to   negative
eigenvalues $\lambda_1$ and $\lambda_2$ at $\xi = + \infty$:
$\vec{a} = b_1 \vec{\psi}_1 \exp(\lambda_1 \xi) + b_2 \vec{\psi}_2
\exp(\lambda_2 \xi)$. Figure~\ref{profile1} shows a typical
solution of Eqs.~(\ref{plane}) for $u=u(\xi)$ and $c=c(\xi)$.

We now perform a linear stability analysis. Consider perturbations in the transverse
direction
\begin{eqnarray}
&&u=u_{0}(\xi)+u_1(\xi)\exp(\gamma t +i k y), \nonumber
\\
&&c=c_{0}(\xi)+c_1(\xi)\exp(\gamma t +i k y)
\label{ansatz}
\end{eqnarray}
and substitute  into Eqs.~(\ref{basicdim}). We have
\begin{eqnarray}
&& \frac{1}{\delta}{u_1}^{\prime\prime} + v {{u_1}^\prime}
+ \left(\frac{2}{m}\,c_0\,u_0 - \frac{k^2}{\delta} - \gamma\right) u_1 + \frac{1}{m}\,{u_0}^2\,c_1 = 0 \,, \nonumber \\
&&{c_1}^{\prime\prime} + v {{c_1}^\prime} - \left({u_0}^2 + k^2 +
\gamma\right) c_1 - 2 c_0\,u_0\,u_1 = 0\,. \label{linear}
\end{eqnarray}
For a fixed value of transverse wave number $k$, we should find
the perturbations $u_1(\xi)$ and $c_1(\xi)$, and the growth rate
$\gamma$. As before, we rewrite Eqs.~(\ref{linear}) as four
coupled first-order differential equations in the form
$(\vec{a}_{lin})^{\prime} = M_{lin} (\vec{a}_{lin})$. There are
two positive eigenvalues of the matrix $M_{lin}$ at $\xi = -
\infty$. We start at $\xi = - \infty$ from the linear combination
of the corresponding eigenvectors, $\vec{a}_{lin} = b_3
\vec{\psi}_3 \exp(\lambda_3 \xi) + b_4 \vec{\psi}_4 \exp(\lambda_4
\xi)$, where one can chose $b_4$ to be unity due to the linearity
of the problem. Performing shooting in two parameters, the growth
rate $\gamma$ and the constant $b_3$, we find the eigenfunctions
$u_1(\xi)$ and $c_1(\xi)$ by demanding that the solution
$\vec{a}_{lin}$ is given at $\xi = + \infty$ by a linear
combination of the two eigenvectors $\vec{\psi}_5$ and
$\vec{\psi}_6$ belonging to negative eigenvalues
$\lambda_5$ and $\lambda_6$, $\vec{a}_{lin} = b_5 \vec{\psi}_5
\exp(\lambda_5 \xi) + b_6 \vec{\psi}_6 \exp(\lambda_6 \xi)$.
Changing the value of transverse wave number $k$, for the fixed
$\delta$ and $m$, we calculate the dispersion curve $\gamma(k)$.

\begin{figure}[ht]
\vspace{-0.4cm}
\centerline{\includegraphics[width=6.5cm,clip=]{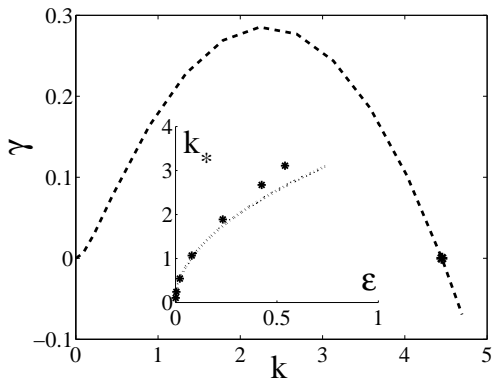}}
\caption{An example of the dispersion curve $\gamma(k)$. The
instability occurs if the ratio of diffusion coefficients $\delta$
exceeds a certain critical value. For $\delta > \delta_{cr}\approx
2.300$, the growth rate $\gamma$ is positive for small $k$, while
for larger $k$, cell diffusion in the transverse direction
stabilizes the instability. The parameters are: $m=0.1$, $\delta =
20$. The inset shows the dependence of the largest unstable wave
number $k_*$ on  $\epsilon = (\delta -
\delta_{cr})/\delta_{cr}$. Numerical simulations are the
asterisks, the asymptote $k_* = (0.36/m)\epsilon^{1/2}$ is 
the dotted line.}
\label{dispersion}
\end{figure}

As was found previously in the context of chemical reactions for
$m=1$ \cite{Horvath}, plane fronts can become transversally
unstable if the ratio of diffusion coefficients $\delta$ exceeds a
certain critical value. Indeed, for $\delta > \delta_{cr}$, the
growth rate $\gamma$ is positive for small $k$, while for larger
$k$, cell diffusion in the transverse direction stabilizes the
instability. We checked that $\delta_{cr} \approx 2.300$, in
agreement with previous results \cite{Horvath,MALEVANETS}.
Figure~\ref{dispersion} shows an example of the dispersion curve for
$\delta > \delta_{cr}$. An inset shows the dependence of the
largest unstable wave number $k_*$ on 
$\epsilon = (\delta - \delta_{cr})/\delta_{cr}$. Numerical
simulations are denoted by asterisks, the asymptote $k_* =
(0.36/m)\epsilon^{1/2}$ is shown by the dashed line.

For a very wide system the instability threshold
$\delta_{cr}$ does not depend on $m$. To see this,  introduce
new dimensionless variables $r = (m/\delta^{1/2})R$, $t =
m^2T$, and $u = U/m$. In this case, $m$ drops out of
the problem. A consequence of
the elimination of $m$  is that one can easily
find the dependence of the velocity,
$v$, and of the wave number, $k$, on $m$: $v =
m^{-1}\delta^{-1/2} V$, $k = m^{-1}\delta^{1/2} K$.  Since
the scaled front velocity, $V$,  and the scaled wave
number, $K$, must be independent of $m$,  $v$
and $k$ are proportional to $m^{-1}$. We will not eliminate the
parameter $m$ from the problem and will work with
Eqs.~(\ref{basicdim}). Different types of cells have different
diffusion coefficients $D_u$ and different proliferation rates
$\alpha$. Therefore, it is convenient that the dependence of the
physical quantities $k_{phys} = k [D_c/(\beta {u_0}^2)]^{-1/2}$
and $v_{phys} = v (D_c \beta {u_0}^2)^{1/2}$ on $D_u$ and $\alpha$
enters only via the dimensionless variables $v$ and $k$.

For a system of finite width we need to form a discrete set of modes from the 
modes of the infinite-width system by imposing $kL=2n\pi$, where $n=1, 2, \dots$ and $L$ is the width of the system. For a small enough system we can `freeze out' the instability if $k_* < 2\pi /L.$ For a spherical tumor $L$ is of the order of the diameter. 
For example,  Figure~\ref{tumor}(a), could correspond to small $k_*$.  

\begin{figure}[ht]
\vspace{-0.4cm}
\centerline{\includegraphics[width=6.5cm,clip=]{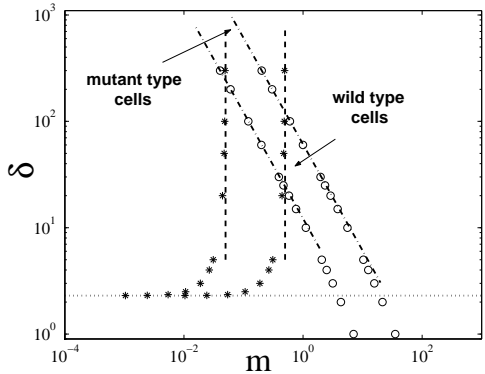}}
\caption{The phase plane $(\delta, m)$ with  two
families of curves: $v = \mbox{const}$ and $k_* = \mbox{const}$,
 calculated from Eqs.~(\ref{plane}) and (\ref{linear}). The
two curves marked by $\circ$'s show $v = \mbox{const}$;  $v = 0.1$ (left), and
 $v = 0.02$ (right) . The two curves marked by $*$'s show $k_* =\mbox{const}$ . 
The left-hand curve  corresponds
to a stronger instability, $k_* = 10$, while the right one
corresponds to a weaker instability, $k_* = 1$. Also shown are the
large $\delta$ asymptotes: $\delta = \mbox{const}/m$ for $v =
\mbox{const}$ curves, and $m = \mbox{const}$ for $k_* =
\mbox{const}$ curves, see text. The instability threshold for an
infinite stripe, $\delta \approx 2.300$, is plotted by the dotted
line. The larger-$\delta$ and smaller-$m$ region correspond to
wild-type cells, while mutant  cells are in the region of
smaller $\delta$ and larger $m$. This suggests that wild-type
cells have a larger diffusion constant, but a smaller
proliferation rate, compared to mutant cells.}
\label{phaseplane}
\end{figure}

We now consider a phase plane of parameters $(\delta,m)$, see
Fig.~\ref{phaseplane}. As mentioned above, the main differences
between the experiments with wild-type and mutant cells are in the
velocity of the front propagation and possible symmetry-breaking.
It means that the region of larger $v$ and smaller $k_*$ in this
phase plane corresponds to wild-type cells, while the region of
smaller $v$ and larger $k_*$ corresponds to mutant cells, see
Fig.~\ref{phaseplane}. 

We consider two families of
curves: the first is of constant front velocity $v =
\mbox{const}$ and the second is of constant largest unstable wave
numbers $k_* = \mbox{const}$. We focus first on $v = \mbox{const}$
curves. It was shown previously, that for $m=1$ and large
$\delta$, $v = 1.219 \delta^{-1}$ \cite{Needham}. Combining this
with the $m$ dependence, one can see that in the $(\delta,m)$
phase plane the curves of constant velocities for large $\delta$
are given by $\delta = \mbox{const}/m$. Then, we consider $k_* =
\mbox{const}$ curves. Our numerical calculations indicate that for
large values of $\delta$, the largest unstable wave number $k_*$
tends to some constant independent of $\delta$; the value of this
constant for $m=1$ is approximately $0.5$. Therefore, for large
$\delta$, $k_* = 0.5/m$, and the curves of constant $k_*$ are
given by $m = \mbox{const}$. Figure~\ref{phaseplane} shows also
these large $\delta$ asymptotes. A typical length scale is
$[D_c/(\beta {u_0}^2)]^{1/2}\sim 0.2 \mbox{cm}$ and a typical
velocity scale $(D_c \beta {u_0}^2)^{1/2} \sim 4\times 10^{-7}
\mbox{cm}/\mbox{s}$. Comparing this with experimental data
\cite{Stein}, one can see that the dimensionless wave number $k_*$
and front velocity $v$ should be of the order of $5$ and $0.1$,
correspondingly.

As one can see on Fig.~\ref{phaseplane}, larger-$\delta$ and
smaller-$m$ region corresponds to wild-type cells, while mutant
type cells correspond to smaller-$\delta$ and larger-$m$ region.
Thus we predict that wild-type cells have a larger diffusion constant
but a smaller proliferation rate than mutant cells in the invasive zone.
Comparing these theoretical predictions with experiments
\cite{Stein}, two points should be taken into account. First,
experiments \cite{Stein} with tumor cells were performed in a
three-dimensional geometry, so the tumor growth can be described
by spherical propagating fronts for wild-type cells. In order to
explain branching patterns of mutant type cells, linear stability
analysis of spherical fronts, rather than plane fronts, should be
performed. In this case, the role of the initial tumor radius should
be analyzed. Second, the basic solutions of our model are plane
fronts which propagate with constant velocity. However, in
experiments, the more dense inner proliferative region grows
slower than the less dense outer invasive region \cite{Stein}.
Probably the tumors are in a  transient regime, and the steady-state behavior will set in later \cite{Khain}. However,  our 
qualitative predictions should hold in three dimensions. Experimental verification of these
features of the growth should be possible. 

In summary, we have considered the growth of GBM tumors. 
\emph{In vitro} experiments \cite{Stein}
showed that the dynamics of growth and resulting patterns are
quite different for wild-type and mutant cells. For the wild-type
the invasive region grows faster, and tumor remains
spherically symmetric. On the other hand, the invasive region
grows slower for the mutant cells, and there are indications
of symmetry-breaking of spherically symmetric growth. We
formulated a simple reaction diffusion model that captures these
experimental findings. Based on our model, we explain
different patterns by different diffusion constants and
proliferation rates of wild-type and mutant cells: wild-type cells
diffuse faster, but have a lower proliferation rate in the invasive zone.
We think that an attempt should be made to test these predictions and relate them to the microscopic biology of the two cell lines.

\begin{acknowledgments} We would like to thank Andy Stein for many useful
conversations and   T. Demuth and M. Berens for experimental results.
Supported by NIH Bioengineering Research Partnership grant R01 CA085139-01A2.
\end{acknowledgments}

\end{document}